\documentstyle[twoside,fleqn,npb,epsfig]{article}

\def\pbi{\,{\rm pb}^{-1}}
\def\tev{\,{\rm TeV}}
\def\gev{\,{\rm GeV}}
\def\epslam{\epsilon / \Lambda^2}

\title{Contact Interactions: Results from ZEUS and a Global Analysis}

\author{A.F.\.Zarnecki\address{Institute of Experimental Physics, 
        Warsaw University, Ho\.za 69, 00-681 Warszawa, Poland}
        (zarnecki@fuw.edu.pl) }

\begin{document}

\begin{abstract}
In a search for signatures of physics processes beyond the Standard Model,
various $eeqq$ vector contact interaction hypotheses have been tested
in the ZEUS experiment.
No significant evidence of a contact interaction signal has been found.  
The analysis is based on NC $e^+p$ DIS data corresponding
to an integrated luminosity of $47.7\pbi$ and results in 95\% CL limits on the
effective mass scales $\Lambda$ ranging from 1.7 to $5\tev$ 
for the different one-parameter contact interaction scenarios considered.

Within the global analysis, including data from other experiments as well,
any contact interactions with mass scale below $2.1\tev$ are  
excluded at 95\% CL.
Combined mass scale limits for specific one-parameter scenarios range
from 5.1 to $18\tev$.
Upper limits on 
possible effects to be observed in future HERA, LEP and Tevatron running
are estimated. 
The total hadronic cross-section at LEP and $e^{-}p$ scattering
cross-section at HERA are strongly constrained by existing data,
whereas large cross-section deviations are still possible for 
Drell-Yan lepton pair production at the Tevatron.
\end{abstract}

\maketitle

\section{INTRODUCTION}
\label{sec-int}

The HERA $ep$ collider has extended the kinematic range available for the study
of deep--inelastic scattering (DIS) by two orders of magnitude to values of
$Q^2$ up to about $50000\gev^2$.
 Measurements in this
domain allow new searches for physics processes beyond the Standard Model (SM)
at characteristic mass scales in the $\tev$ range. 
The recent analyses were stimulated
in part by an excess of events over the SM expectation for $Q^2 > 20000\gev^2$
reported in 1997 by the ZEUS \cite{zeus97} and H1 \cite{h1_97}
collaborations, for which electron-quark contact interactions (CI)
have been suggested as possible explanations.


\begin{table*}[t]
\caption{Coupling structure and the 95\% CL limits on the effective 
mass scales for different one-parameter contact interaction models.
Results from the ZEUS analysis based on high-$Q^{2}$ $e^{+}p$ NC DIS data
are compared to results based on the combined electron/muon NC data
(global analysis not assuming SU(2) invariance) and those corresponding to 
all available data (global analysis with SU(2) universality).
}
\label{tab-limits}
\setlength{\tabcolsep}{1mm}
\begin{tabular*}{\textwidth}%
    {@{}l@{\hspace{5mm}}cccc@{\extracolsep{\fill}}rrrrrr}
\hline
  & & & & & \multicolumn{2}{c}{\hspace*{-1cm}ZEUS analysis\hspace*{-1cm}}  
          & \multicolumn{4}{c}{Global analysis}  \\ \cline{8-11}
  Model   & \multicolumn{4}{c}{Couplings}  
          & \multicolumn{2}{c}{ \hspace*{-1cm}
                               $e^{+}p$ NC DIS data \hspace*{-1cm}}  
          & \multicolumn{2}{c}{$e/\mu$ NC data}
          & \multicolumn{2}{c}{All data} \\ 
            \cline{2-5} \cline{6-7} \cline{8-9} \cline{10-11}
          & $\eta^{eq}_{LL}$    & $\eta^{eq}_{LR}$
          & $\eta^{eq}_{RL}$    & $\eta^{eq}_{RR}$ 
          & $\Lambda_{min}^{-}$ & [TeV]~~ $\Lambda_{min}^{+}$
          & $\Lambda_{min}^{-}$ & [TeV]~~ $\Lambda_{min}^{+}$ 
          & $\Lambda_{min}^{-}$ & [TeV]~~ $\Lambda_{min}^{+}$  \\ \hline

 %
 VV & + & + & + & + & 5.0 & 4.7 &  9.8 & 10.7 &  9.6 & 11.4 \\
 AA & + & --& --& + & 2.6 & 2.5 & 10.5 & 10.1 &  9.9 & 11.1 \\
 VA & + & --& + & --& 3.7 & 2.6 &  6.6 &  6.2 &  6.3 &  8.0 \\
 %
 X1 & + & --&   &   & 2.8 & 1.8 &  8.7 &  8.1 &  8.1 &  9.5 \\
 X2 & + &   & + &   & 3.1 & 3.4 &  8.2 &  8.4 &  7.8 &  9.6 \\
 X3 & + &   &   & + & 2.8 & 2.9 &  9.9 & 10.2 &  9.5 & 11.1 \\
 X4 &   & + & + &   & 4.3 & 4.0 &  5.7 &  5.2 &  6.0 &  5.4 \\
 X5 &   & + &   & + & 3.3 & 3.5 &  5.9 &  6.4 &  6.2 &  6.4 \\
 X6 &   &   & + & --& 1.7 & 2.8 &  6.2 &  5.8 &  6.2 &  5.8 \\
 \hline
 %
 U1 & + & --&   &   & 2.6 & 2.0 &  6.4 &  7.7 &  7.9 & 17.0 \\
 U2 & + &   & + &   & 3.9 & 4.0 &  6.9 &  9.1 &      &      \\
 U3 & + &   &   & + & 3.5 & 3.7 &  8.5 & 11.7 &  8.6 & 18.2 \\
 U4 &   & + & + &   & 4.8 & 4.4 &  5.1 &  5.5 &      &      \\
 U5 &   & + &   & + & 4.2 & 4.0 &  6.4 &  8.8 &  7.1 &  8.8 \\
 U6 &   &   & + & --& 1.8 & 2.4 &  7.0 &  5.6 &      &       \\ 
 \hline
\end{tabular*}
\end{table*}

\section{CONTACT INTERACTIONS}
\label{sec-ci}

Four-fermion contact interactions are an effective theory, which 
allows us to describe, in the most general way, possible low energy 
effects  coming from ``new physics'' at much higher energy scales. This 
includes the possible existence of second-generation heavy weak bosons, 
leptoquarks as well as electron and quark compositeness \cite{cidef,cihera}.
As strong limits beyond the HERA
sensitivity have already been placed on the scalar and 
tensor terms \cite{cihera},
only the vector $eeqq$ contact interactions
are considered in this study.
They can be represented 
as additional term in the Standard Model Lagrangian \cite{cihera}:
\begin{eqnarray}
L_{CI} & = & \epsilon \frac{g^2}{\Lambda^2}
           \sum_{i,j=L,R} \eta^{eq}_{ij} (\bar{e}_{i} \gamma^{\mu} e_{i} )
              (\bar{q}_{j} \gamma_{\mu} q_{j}) 
\end{eqnarray}
where the sum runs over electron and quark helicities,
$\epsilon$ is the overall sign of the CI Lagrangian,
$g$ is the coupling, and $\Lambda$ is the effective mass scale.
Helicity and flavour structure of contact interactions is described by 
set of parameters $\eta^{eq}_{ij}$.
Since $g$ and $\Lambda$ always enter in the
combination $g^2/\Lambda^2$, we fix the coupling by adopting
the convention $g^2=4\pi$.
In the ZEUS analysis 30 specific CI scenarios are considered.
Assumed relations between different couplings are
listed in Table \ref{tab-limits}.
Each line in the table represents 
two scenarios, one for $\epsilon=+1$ and one for $\epsilon=-1$.
For the models VV, AA, VA and X1 to X6 all quark flavours are assumed 
to have the same contact interaction couplings and
each of the $\eta_{ij}^{eq}$ is either zero or $\pm 1$.
For the U1 to U6 models only couplings of up-type quarks ($u$ and $c$)
are considered.

The global analysis combining data from different experiments 
(see sections \ref{sec-global} and \ref{sec-pred})  also 
takes into account three less constrained models, 
in which different couplings can vary independently.
    The {\bf General Model} assumes that contact interactions couple only 
    electrons to $u$ and $d$ quarks (8 independent couplings). 
    All other couplings (for $s, c, b, t, \mu , \tau$) are 
    assumed to vanish.
    The model with {\bf Family Universality} assumes lepton universality
    ($e$=$\mu$) and quark family universality ($u$=$c$ and 
    $d$=$s$=$b$). There are also 8 independent couplings.
    In a model assuming ${\bf SU(2)_{L} \times U(1)_{Y}}$ gauge invariance,
    the number of free model parameters is reduced from 8 to 7
    ($\eta^{eu}_{RL}$=$\eta^{ed}_{RL}$). In this model the $eeqq$ 
    contact interaction couplings can be also related to 
    $\nu \nu qq$  and $e \nu q q'$ couplings \cite{ci_su2}.

\section{ZEUS ANALYSIS}
\label{sec-zeus}

This analysis \cite{zeus_ci} is based on  $47.7\pbi$ of
NC $e^+p$ DIS data collected by the ZEUS experiment in 1994-97.
Monte Carlo simulation, event selection, kinematic reconstruction, 
and assessment of systematic effects is that of
the NC DIS analysis described in \cite{zeus_nc}. 
The event sample used in the CI analysis is limited to 
$0.04\!<\!x\!<\!0.95$, $0.04\!<\!y\!<\!0.95$ and $Q^{2}>500\gev^{2}$.

\begin{figure}[t]
\epsfig{figure=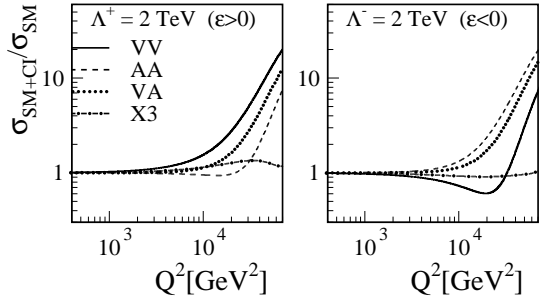,width=7.5cm,clip=} 
\vspace{-1.0cm}
\caption{Examples of the relative influence of a CI on the $e^{+}p$ NC DIS 
cross-section $d\sigma/dQ^{2}$.}
\label{fig-model}
\end{figure}

A cross-section increase at highest $Q^{2}$, corresponding to 
the direct ``new physics'' contribution, is expected for 
most CI scenarios, as shown in Figure \ref{fig-model}.
At intermediate $Q^{2}$ a moderate increase or decrease 
due to CI-SM interference terms is possible.
As the helicity structure of new interactions can be different 
from that of the Standard Model, also the differential cross-section
$d\sigma / dx$ (for fixed $Q^{2}$) is modified.
Sensitivity to many CI scenarios
is significantly improved by considering
the two-dimensional event distribution.

The ZEUS CI analysis compares the distributions of the measured kinematic 
variables with the corresponding distributions from a MC simulation 
of $e^+p\to e^+X$ events reweighted to simulate the CI scenarios.
An unbinned log--likelihood technique is used to calculate
$L(\epslam)$ from the individual kinematic event coordinates $(x_i,y_i)$:
\begin{equation}
  L(\epslam)=-\sum_{i\in{\rm data}}\log{p(x_i,y_i;\epslam)} \; ,
\end{equation}
where the sum runs over all events in the selected data sample and 
$p(x_i,y_i;\epslam)$ is the probability that an event $i$
observed at  $(x_i , y_i)$ results from the model described by coupling 
$\epslam$. $L$ tests the shape of the $(x,y)$--distribution but
is independent of its absolute normalisation.

The best estimates, $\Lambda^{\pm}_\circ$, for the different CI scenarios 
are given by the positions of the respective minima of $L(\epslam)$ for
$\epsilon$=$\pm 1$.
All results are consistent with the Standard Model,
the probability that the observed values of $\Lambda^{\pm}_{\circ}$
result from the Standard Model does not fall below 16\%.
The $95\%$ C.L. lower limits $\Lambda_{min}$ on the effective
mass scale $\Lambda$ are defined as the mass
scales for which MC experiments have a 95\% chance to result in 
$\Lambda_{\circ}$ values smaller than that observed in data.
The lower limits on $\Lambda$ ($\Lambda^\pm_{min}$ for $\epsilon$=$\pm 1$)
are summarized in Table \ref{tab-limits}.
The $\Lambda$ limits range from 1.7 to $5\tev$. 

\section{GLOBAL ANALYSIS}
\label{sec-global}

The global analysis \cite{my_ci} of $eeqq$ contact interactions
combines relevant data from different experiments:
ZEUS and H1 high-$Q^{2}$ NC DIS results;
Tevatron data on high-mass Drell-Yan lepton pair production;
LEP2 results on the hadronic cross-section 
            $\sigma ( e^{+}e^{-} \rightarrow q \bar{q} (\gamma) )$,
 the heavy quark production ratios $R_{b}$ and $R_{c}$, and the
 forward-backward asymmetries  $A^{b}_{FB}$, $A^{c}_{FB}$, $A^{uds}_{FB}$;
data from low-energy $eN$ and $\mu N$ scattering
and from atomic parity violation (APV) measurements.

For models assuming $SU(2)_{L} \times U(1)_{Y}$ universality,
additional constraints come from 
HERA $e^{+}p$  CC DIS results,
data on $\nu N$ scattering,
unitarity  of the CKM matrix and  electron-muon universality.

The combined data are consistent with the Standard Model predictions.
The mass scale limits  $\Lambda_{min}^{-}$ and $\Lambda_{min}^{+}$
obtained from fitting one-parameter models to all available data
are summarized in Table \ref{tab-limits}.
For models not assuming $SU(2)_{L} \times U(1)_{Y}$  universality
(only $e$/$\mu$ NC data used) the mass limits range
from  5.1 to   $11.7\tev$.
With $SU(2)_{L} \times U(1)_{Y}$  universality (using also $\nu N$ and CC data)
the limits extend up to about $18\tev$.

Limits for single couplings derived in multi-parameter models
(of Section \ref{sec-ci})
are weaker than in the case of one-parameter models, as no relation
between separate couplings is assumed.
The mass limits obtained for the general model range 
from 2.1 to $5.1\tev$.
All limits improve significantly  
and reach  3.5 to $7.8\tev$ 
for the SU(2) model with family universality.

Taking into account possible correlations between couplings,
any contact interaction with a mass scale below $2.1\tev$
($3.1\tev$ when SU(2) universality is assumed) is excluded at 95\% CL.

\section{PREDICTIONS}
\label{sec-pred}

\begin{figure}[t]
\epsfig{figure=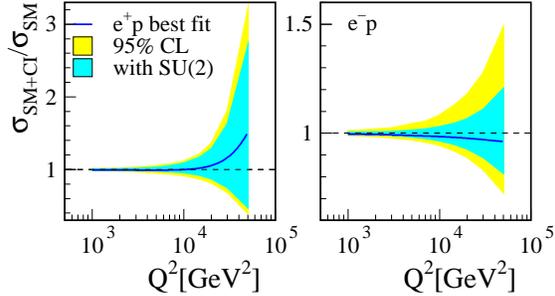,width=7.5cm,clip=} 
\vspace{-1.0cm}
\caption{The 95\% CL limit band on the ratio of predicted
to the Standard Model cross-section for  $e^{+}p$ and $e^{-}p$ NC DIS
scattering at HERA.}
\label{fig-predict}
\end{figure}

Likelihood function for the possible cross-section deviations
from the Standard Model predictions is calculated as the weighted
average over all contributing CI coupling combinations \cite{my_ci}.
The results for HERA, in terms of the 95\% C.L. limit bands on the
ratio of predicted and the Standard Model cross-sections as a function
of $Q^{2}$, are shown  in Figure \ref{fig-predict},
for the general model and the SU(2) model with family universality.

The allowed increase in the integrated $e^{+}p$ NC DIS cross-section 
for $Q^{2} \; >$ 15,000 GeV$^{2}$
is about 40\% for the general model and about 30\% for the SU(2) model.
In order to reach the level of statistical precision
which would allow to confirm a possible discrepancy of this 
size, the HERA experiments would have to collect $e^{+}p$
integrated luminosities of the order of 100-200$\pbi$ 
(depending on the model). 
This will be possible after the HERA upgrade planned for year 2000.

Constraints on possible deviations from the Standard Model 
predictions are much stronger in the case of $e^{-}p$ NC DIS.
For the general model
deviations larger than about 20\% are excluded for $Q^{2} >$15,000 GeV$^{2}$,
whereas for the SU(2) model with family universality the limit goes down
to about 7\%. 
In such a case it would be very hard to detect contact interactions 
in future HERA $e^{-}p$ running.
However, for scattering with 60\% longitudinal $e^{-}_{R}$ polarisation, 
the maximum allowed deviations increase to 28\% and 19\%, respectively,
and significant effects could be observed already for 
integrated luminosities of the order of 120$\pbi$.

For the hadronic cross-section at LEP, for 
\mbox{$\sqrt{s}\!\sim\!200 \gev$},
the possible deviations from the Standard Model are
only about 8\%. 
However, significant deviations
are still possible for the heavy quark production ratios 
$R_{c}$ and $R_{b}$, and for the forward-backward asymmetries 
$A^{c}_{FB}$ and $A^{b}_{FB}$.
Significant cross-section deviations will be possible in 
the Next Linear Collider (NLC), for  $\sqrt{s}\! > \! 300 \gev$.
The largest cross-section deviations from the Standard Model predictions are 
still allowed at the Tevatron. For Drell-Yan lepton pair production,
deviations of the cross-section at  $M_{ll}$=500 GeV up to a factor of 5 
are still not excluded. 

\begin{figure}[t]
\epsfig{figure=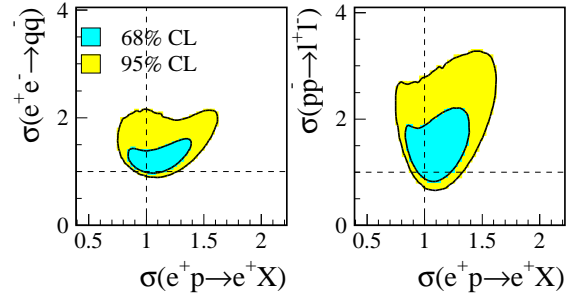,width=7.5cm,clip=} 
\vspace{-1.0cm}
\caption{The 68\% and 95\% CL contours for the possible deviation from
the Standard Model predictions, for $e^{+}p$ NC DIS
cross-section at HERA, at $Q^{2}$ = 30,000 GeV$^{2}$, total
$e^{+}e^{-} \rightarrow q\bar{q}$ cross-section at $\sqrt{s}$ = 400 GeV
and Drell-Yan lepton pair production cross-section at
the Tevatron, at $M_{ll}$ = 500 GeV, as indicated on the plot. 
The limits are calculated using the SU(2) contact interaction
model with family universality.}
\label{fig-rel}
\end{figure}
In Figure \ref{fig-rel} relations between possible cross-section deviations
at HERA, NLC and the Tevatron are presented. There are no clear correlations
between different experiments.
All experiments should continue to analyse their data in terms of
possible new electron-quark interactions, as constraints resulting from
different processes are, to large extent, complementary.

\end{document}